\documentclass{aa}  
\usepackage{graphicx}
\usepackage{txfonts}
\usepackage{xcolor}
\usepackage{hyperref}
\hypersetup{colorlinks, linkcolor=blue, citecolor=blue, urlcolor=blue}
\usepackage{amsmath}
\usepackage{ulem}
\usepackage{soul}
\usepackage{enumitem}

\newcommand{\muvec}{\mbox{\boldmath $\mu$}}

\newcommand{\te}{t_{\rm E}}
\newcommand{\thetae}{\theta_{\rm E}}

\newcommand{\pie}{\pi_{\rm E}}

\newcommand{\dl}{D_{\rm L}}
\newcommand{\ds}{D_{\rm S}}

\definecolor{brown}{rgb}{0.59, 0.29, 0.0}
\definecolor{darkgreen}{rgb}{0.0, 0.42, 0.24}
\definecolor{darkblue}{rgb}{0.01, 0.31, 0.59}
\definecolor{darkblue}{rgb}{0.0, 0.25, 0.42}
\definecolor{blue}{rgb}{0.0,0.0,1.0}
\definecolor{green}{rgb}{0.0,1.0,0.0}

\def\eqalign#1{\null\,\vcenter{\openup\jot
        \ialign{\strut\hfil$\displaystyle{##}$&$
        \displaystyle{{}##}$\hfil \crcr#1\crcr}}\,}

\begin{document}

\title{KMT-2021-BLG-1122L: The first microlensing triple stellar system}
\titlerunning{KMT-2021-BLG-1122L: The first microlensing triple stellar system}

\author{
     Cheongho~Han\inst{01} 
\and Youn~Kil~Jung\inst{02} 
\and Andrew~Gould\inst{03,04}      
\and Doeon~Kim\inst{01}
\and Chung-Uk~Lee\inst{02} 
\\
(Leading authors)\\
and \\
     Michael~D.~Albrow\inst{05}   
\and Sun-Ju~Chung\inst{02}      
\and Kyu-Ha~Hwang\inst{02} 
\and Hyoun-Woo~Kim\inst{02} 
\and Yoon-Hyun~Ryu\inst{02} 
\and In-Gu~Shin\inst{06} 
\and Yossi~Shvartzvald\inst{07}   
\and Hongjing~Yang\inst{08}     
\and Jennifer~C.~Yee\inst{06}   
\and Weicheng~Zang\inst{06,08}     
\and Sang-Mok~Cha\inst{02,09} 
\and Dong-Jin~Kim\inst{02} 
\and Seung-Lee~Kim\inst{02,10} 
\and Dong-Joo~Lee\inst{02} 
\and Yongseok~Lee\inst{02,09} 
\and Byeong-Gon~Park\inst{02,10} 
\and Richard~W.~Pogge\inst{04}
\\
(The KMTNet Collaboration)\\
}

\institute{
     Department of Physics, Chungbuk National University, Cheongju 28644, Republic of Korea  \\ \email{cheongho@astroph.chungbuk.ac.kr}    
\and Korea Astronomy and Space Science Institute, Daejon 34055, Republic of Korea                                                          
\and Max Planck Institute for Astronomy, K\"onigstuhl 17, D-69117 Heidelberg, Germany                                                      
\and Department of Astronomy, The Ohio State University, 140 W. 18th Ave., Columbus, OH 43210, USA                                         
\and University of Canterbury, Department of Physics and Astronomy, Private Bag 4800, Christchurch 8020, New Zealand                       
\and Center for Astrophysics $|$ Harvard \& Smithsonian 60 Garden St., Cambridge, MA 02138, USA                                            
\and Department of Particle Physics and Astrophysics, Weizmann Institute of Science, Rehovot 76100, Israel                                 
\and Department of Astronomy and Tsinghua Centre for Astrophysics, Tsinghua University, Beijing 100084, China                              
\and School of Space Research, Kyung Hee University, Yongin, Kyeonggi 17104, Republic of Korea                                             
\and Korea University of Science and Technology, 217 Gajeong-ro, Yuseong-gu, Daejeon, 34113, Republic of Korea                             
}
\date{Received ; accepted}

\abstract
{}
{
We systematically inspect the microlensing data acquired by the KMTNet survey during the
previous seasons in order to find anomalous lensing events for which the anomalies in 
the lensing light curves cannot be explained by the usual binary-lens or binary-source 
interpretations.
}
{
From the inspection, we find that interpreting the three lensing events OGLE-2018-BLG-0584, 
KMT-2018-BLG-2119, and KMT-2021-BLG-1122 requires four-body (lens+source) models, in which 
either both the lens and source are binaries (2L2S event) or the lens is a triple system 
(3L1S event).  Following the analyses of the 2L2S events presented in \citet{Han2023}, 
here we present the 3L1S analysis of the KMT-2021-BLG-1122.
}
{
It is found that the lens of the event KMT-2021-BLG-1122 is composed of three masses, in 
which the projected separations (normalized to the angular Einstein radius) and mass ratios
between the lens companions and the primary are $(s_2, q_2)\sim (1.4, 0.53)$ and $(s_3, q_3)
\sim (1.6, 0.24)$. By conducting a Bayesian analysis, we estimate that the masses of the 
individual lens components are $(M_1, M_2, M_3)\sim (0.47\,M_\odot, 0.24\,M_\odot, 
0.11\,M_\odot)$.  The companions are separated in projection from the primary by 
$(a_{\perp,2}, a_{\perp,3})\sim (3.5, 4.0)$~AU.  The lens of KMT-2018-BLG-2119 is the 
first triple stellar system detected via microlensing.
}
{}


\maketitle

\section{Introduction}\label{sec:one}

Light curves of gravitational microlensing events often exhibit deviations from the smooth 
and symmetric form of the event involved with a single lens mass and a single source star 
\citep{Paczynski1986}.  The most common cause of such deviations is the binarity of the lens, 
2L1S event \citep{Mao1991}, or the source, 1L2S event \citep{Griest1993, Han1997}. Anomalous 
lensing events comprise about 10\% of more than 3000 lensing events that are being annually
detected from the three currently working lensing surveys conducted by the Optical Gravitational 
Lensing Experiment \citep[OGLE:][]{Udalski2015}, Microlensing Observations in Astrophysics 
\citep[MOA:][]{Bond2001}, and Korea Microlensing Telescope Network \citep[KMTNet:][]{Kim2016} 
groups.  The prime goal of these survey experiments is finding extrasolar planets, which are 
detected via anomalous signals in lensing light curves.  Planetary microlensing signals appear 
in various shapes and often confused with anomalies produced by other causes.  In order to sort 
out planetary signals from those of other origins, therefore, it is essential to analyze all 
anomalous events.

In the current microlensing experiments, anomalous events are being analyzed almost in real 
time with the progress of events. This became possible with the implementation of early warning 
systems: the OGLE early warning system \citep{Udalski1994}, MOA real-time analysis system 
\citep{Bond2001}, and KMTNet EventFinder AlertFinder \citep{Kim2018b} system.\footnote{In 
addition, the KMTNet EventFinder system \citep{Kim2018a}  carries out end-of-season analyses 
that recover many events that were missed for one reason or another by AlertFinder \citep{Kim2018b} 
system.} These systems enable one to detect lensing events in their early phases and to promptly 
identify anomalies appearing in lensing light curves. Analyses of the events with identified 
anomalies are conducted by multiple modelers of the individual survey groups as the anomalies 
proceed, and models of the events found from these analyses are circulated or posted on web 
pages\footnote{For example, the model page of lensing events maintained by Cheongho Han 
({\tt http://astroph.chungbuk.ac.kr/$\sim$cheongho}).} to inform researchers in the microlensing 
community of the nature of the anomalies.  In most cases, anomalous events are interpreted with 
a binary-lens (2L1S) or a binary-source (1L2S) model.

For a minor fraction of anomalous events, it is known that the anomalies cannot be explained 
by the usual 2L1S or 1L2S interpretation. \citet{Han2023}, hereafter Paper~I, conducted a 
systematic investigation of events found from the KMTNet survey during the previous seasons 
in search for anomalous events for which no plausible models had been presented. From this 
investigation, they found that the anomalies appearing in the light curves of the events 
OGLE-2018-BLG-0584 and KMT-2018-BLG-2119 required a four-body (lens+source) model, in which 
both the lens and source are binaries (2L2S event). In this paper, we present the analysis 
of another four-body lensing event KMT-2021-BLG-1122. Unlike the two 2L2S events presented 
in Paper~I, the lens system of KMT-2021-BLG-1122 is composed of three masses (3L1S).

We present the analysis of the event KMT-2021-BLG-1122 according to the following organization.
In Sect.~\ref{sec:two}, we depict the observations conducted to obtain the photometric data 
of the event, and describe the characteristic features of the anomaly appearing in the light 
curve of the event.  In Sect.~\ref{sec:three},  we begin by describing the lensing parameters 
used in the modeling conducted under various configurations of the lens and source system, 
and we then describe detail of the analyses conducted under the individual configurations 
in the following subsections: 2L1S model in Sect.~\ref{sec:three-one}, 2L2S model in 
Sect.~\ref{sec:three-two}, and 3L1S model in Sect.~\ref{sec:three-three}.  In 
Sect.~\ref{sec:four}, we specify the source star of the event and estimate the angular 
Einstein radius of the lens system.  In Sect.~\ref{sec:five}, we estimate the physical 
parameters including the masses of the individual lens components, the distance to the lens 
system, and the projected separations among the lens components.  We summarize the results 
from the analysis and conclude in Sect.~\ref{sec:six}.

\section{Lensing light curve and observations}\label{sec:two}

The lensing event KMT-2021-BLG-1122 was detected solely by the KMTNet survey. The event 
occurred on a faint source star, with a baseline magnitude of $I_{\rm base}= 20.36$, 
located toward the Galactic bulge field with equatorial coordinates $({\rm RA}, 
{\rm DEC})= $ (17:35:51.10, $-$28:26:47.22), which correspond to the Galactic coordinates 
$(l, b)=(-0^\circ\hskip-2pt .720,  2^\circ\hskip-2pt .076)$.  The source lies in the KMTNet 
BLG14 field, toward which observations were conducted with a 1~hr cadence. The event was 
identified by the KMTNet AlertFinder system and an alert was issued on 2021 June 3, which 
corresponds to the abridged Heliocentric Julian Date of ${\rm HJD}^\prime \equiv {\rm HJD}
-2450000 \sim 9368.5$.

Observations of the event were carried out with the use of the three 1.6~m telescopes operated 
by the KMTNet group. The telescopes are distributed in the three continents of the Southern
Hemisphere for continuous coverage of lensing events, and the sites of the individual telescopes
are the Siding Spring Observatory in Australia (KMTA), the Cerro Tololo Interamerican Observatory
in Chile (KMTC), and the South African Astronomical Observatory in South Africa (KMTS). The
camera mounted on each telescope is composed of four $9{\rm k}\times 9{\rm k}$ CCD chips,
which in combination yield a 4~deg$^2$ field of view.

Images containing the source were mainly taken in the $I$ band, and about one tenth of images
were obtained in the $V$ band. Reductions of the images and photometry of the event were done
using the KMTNet pipeline constructed based on the pySIS code of \citet{Albrow2009}, and 
additional photometry was carried out for a subset of the KMTC data set using the pyDIA code 
of \citet{Albrow2017}.  The analysis of the event was done based on the $I$-band light curve 
constructed from the pySIS reduction, and the pyDIA photometry data were used for the source 
color measurement. We discuss the detailed procedure of the source color measurement in 
Sect.~\ref{sec:four}. According to recipe described in \citet{Yee2012}, we readjusted the
error bars of the data determined by the automated pipeline in order that the errors were 
consistent with the scatter of data and $\chi^2$ per degree of freedom (dof) for each data set 
became unity.

\begin{figure}[t]
\includegraphics[width=\columnwidth]{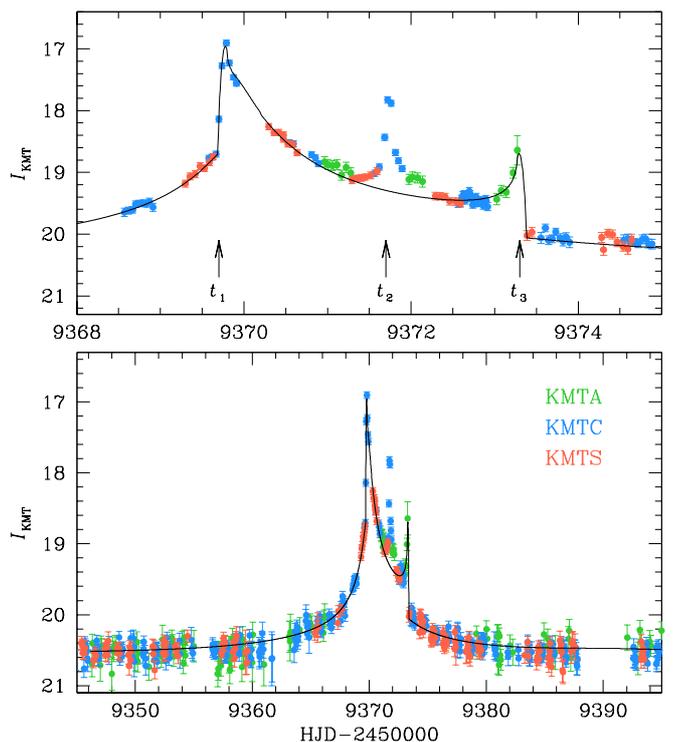}
\caption{
Light curve of KMT-2021-BLG-1122 constructed with the combined data obtained from the three 
KMTNet telescopes (KMTA, KMTC, and KMTS). The curve drawn over the data points is the best-fit 
model found under the 2L1S interpretation.  The lower panel displays the whole view, and the 
upper panel shows the enlarged view of the anomaly region.  The epochs marked by $t_1$, $t_2$, 
and $t_3$ indicate the three major anomaly features.
}
\label{fig:one}
\end{figure}

\begin{figure*}[t]
\centering
\includegraphics[width=12.9cm]{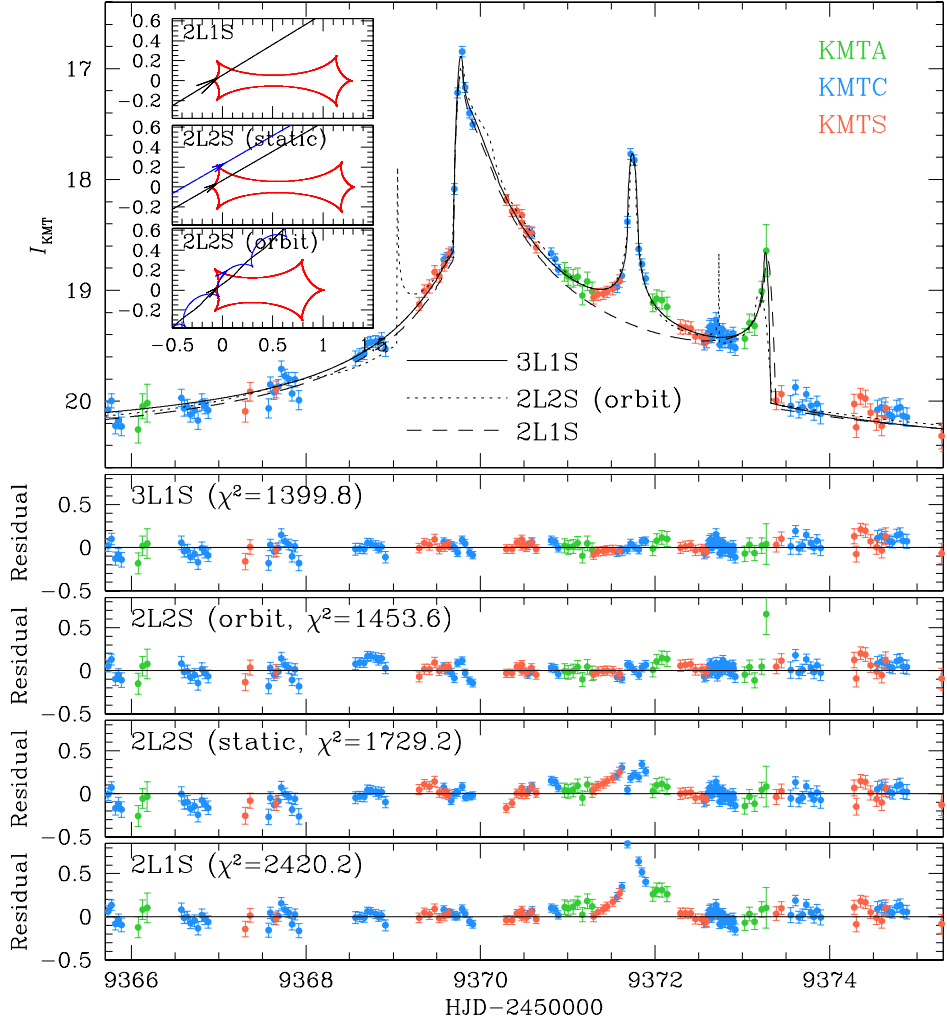}
\caption{
Enlarged view around the central anomaly region of the light curve.  Three model curves 
obtained under the 3L1S, 2L2S, and 2L1S lens-system configurations are drawn over the data 
points, and the three lower panels show the residuals from the individual models. The two 
insets in the top panel show the lens-system configurations of the 2L1S and 2L2S models. 
In each inset, the red figure is the caustic and the line with an arrow represents the 
source trajectory. For the 2L2S configuration, there are two source trajectories, in which 
the black and blue ones represent the trajectories of the primary and secondary source stars, 
respectively.  
}
\label{fig:two}
\end{figure*}

Figure~\ref{fig:one} shows the lensing light curve of KMT-2021-BLG-1122 constructed with the 
combined $I$-band pySIS photometry data obtained from the three KMTNet telescopes.  It shows 
a strong anomaly relative to a 1L1S form, and the anomaly displays a complex pattern that is 
characterized by the three strong features appearing at ${\rm HJD}^\prime \sim 9369.7$ ($t_1$), 
$\sim 9371.7$ ($t_2$), and $\sim 9373.3$ ($t_3$).  From the characteristic profiles described 
in \citet{Schneider1986}, the anomalies at around $t_1$ and $t_3$ appear to be a pair of 
caustic-crossing features produced when a source enters and exits a caustic, respectively. 
On the other hand, the profile at around $t_2$, which is approximately symmetric with respect 
to the peak, appears to be produced when a source approaches close to or crosses over the  
cusp of a caustic \citep{Schneider1987}.

In general, the light curve profile of a 2L1S event in the region between a pair caustic-crossing 
spikes exhibits a "U"-shape pattern.  Occasionally, deviations from this U-shape pattern can 
arise if a source sweeps one fold of a caustic, for example, OGLE-2016-BLG-0890 \citep{Han2022-0890}, 
but, in this case, the resulting deviation is smooth and thus differs from the sharp pattern 
of the observed anomaly at around $t_2$.  This suggests that a different explanation is needed 
to explain the origin of the anomaly.

\section{Interpreting the anomaly}\label{sec:three}

In order to explain the anomalies in the lensing light curve, we conducted a series of modeling
under various interpretations of the lens-system configuration. We tested three models under the
2L1S, 2L2S, and 3L1S configurations.  We exclude 1L2S and 1L3S models because the observed light 
curve exhibits obvious caustic-crossing features, that is, those around $t_1$ and $t_3$, while 
1L2S and 1L3S events do not involve caustics.\footnote{There has been only one 1L3S microlensing 
event published to date: OGLE-2015-BLG-1459 \citep{Hwang2018}.}

In the modeling under each tested interpretation, we search for a lensing solution, representing 
a set of lensing parameters that characterize the observed lensing light curve. The basic lensing 
parameters of a 1L1S event include $(t_0, u_0, \te)$, which denote the time of the closest source 
approach to the lens, the source-lens separation (impact parameter) at $t_0$, and the Einstein 
time scale, respectively. The Einstein time scale represents the time required for the source to 
transit the angular Einstein radius of the lens, $\thetae$, and the length of $u_0$ is scaled 
to $\thetae$.

A 2L1S system corresponds to the case in which the lens contains an extra component compared
to the 1L1S system. Consideration of the extra lens component requires one to include additional
parameters in modeling. These additional parameters are $(s, q, \alpha, \rho)$,  and they denote 
the projected separation (normalized to $\thetae$), and mass ratio between the lens components 
($M_1$ and $M_2$), the angle (source trajectory angle) between the relative lens-source proper 
motion, $\muvec$, and the axis connecting $M_1$ and $M_2$, and the ratio between the angular 
radius of the source, $\theta_*$, to the Einstein radius, that is, $\rho=\theta_*/\thetae$ 
(normalized source radius), respectively. The normalized source radius is included in modeling 
because a 2L1S event usually involves caustic crossings, during which lensing magnifications 
are affected by finite-source effects \citep{Bennett1996}.  For the computation of finite-source 
magnifications, we utilize the map-making method of \citet{Dong2006}.

For the modeling of a 2L2S system, it is required to include extra parameters in addition to 
those of the 2L1S model in order to describe an extra source. These extra parameters include 
$(t_{0,2}, u_{0,2}, \rho_2, q_F)$, where the first three denote the closest approach time, 
impact parameter, and normalized radius of the source companion, and the last parameter 
indicates the flux ratio between the companion ($S_2$) and primary ($S_1$) of the source. In 
the 2L2S modeling, we use the notations $(t_{0,1}, u_{0,1}, \rho_1)$ to denote the parameters 
related to $S_1$.

A 3L1S system has an extra lens component ($M_3$) compared to the 2L1S system, and this 
requires one to include additional parameters in modeling. These parameters include $(s_3, 
q_3, \psi)$, which denote the separation and mass ratio between $M_1$ and $M_3$, and the 
orientation angle of $M_3$ as measured from the $M_1$--$M_2$ axis with a center at the 
position of $M_1$, respectively.  In order to distinguish the parameters describing $M_2$ 
from those describing $M_3$, we use the notations $(s_2, q_2)$ to denote the separation 
and mass ratio of $M_2$.  The lensing parameters for the 1L1S, 2L1S, 2L2S, and 3L1S models 
are summarized in Table~1 of Paper~I.

\subsection{2L1S interpretation}\label{sec:three-one}

We started with modeling the light curve under the 2L1S interpretation of the anomalies. In 
this modeling, we searched for the binary parameters $s$ and $q$ via a grid approach, while 
we seek the other parameters via a downhill approach using a Markov Chain Monte Carlo (MCMC) 
algorithm.  We first constructed a $\Delta\chi^2$ map on the plane of the grid parameters, 
identified local minima, and then refined the individual minima by allowing all lensing 
parameters to vary. From the 2L1S modeling, we found no lensing solution that could describe 
all the anomaly features, and this explains why the event had been left without a plausible 
model.

Although the light curve could not be explained by a 2L1S model, we found a solution that could
approximately describe the caustic-crossing features at around $t_1$ and $t_3$. The model curve 
of this solution, found from modeling the light curve with the exclusion of the data around $t_2$, 
is drawn over the data points in Figures~\ref{fig:one} and \ref{fig:two}, and the corresponding 
configuration of the lens system is presented in the upper inset of the top panel of 
Figure~\ref{fig:two}.  The binary lensing parameters of the solution are $(s, q)\sim (1.69, 
0.43)$, which results in a single resonant caustic elongated along the binary axis. According 
to this solution, the source entered the caustic by crossing the caustic fold lying just above 
the left side on-axis cusp, passed through the inner region of the caustic, and then exited the 
caustic by crossing the upper fold of the caustic. The caustic entrance and exit resulted in 
the anomaly features matching well the observed ones around $t_1$ and $t_3$, respectively.

\begin{table}[t]
\small
\caption{Lensing parameters of the 2L2S solutions\label{table:one}}
\begin{tabular*}{\columnwidth}{@{\extracolsep{\fill}}llccc}
\hline\hline
\multicolumn{1}{c}{Parameter}    &
\multicolumn{1}{c}{Static}       &
\multicolumn{1}{c}{Orbital}      \\
\hline
$\chi^2$/dof                   &  1729.1/1426                    &   1453.6/1429               \\
$t_{0,1}$ (HJD$^\prime$)       &  $9371.208 \pm  0.026  $   &  $9370.681 \pm  0.007  $    \\
$u_{0,2}$                      &  $   0.057 \pm  0.002  $   &  $0.044 \pm  0.001     $    \\
$t_{0,2}$ (HJD$^\prime$)       &  $9370.810 \pm  0.024  $   &  --                         \\
$u_{0,2}$                      &  $   0.194 \pm  0.004  $   &  --                         \\
$\te$ (days)                   &  $  14.36  \pm  0.31   $   &  $14.87  \pm  0.11     $    \\
$s$                            &  $   1.723 \pm  0.020  $   &  $1.447 \pm  0.006     $    \\
$q$                            &  $   0.649 \pm  0.043  $   &  $0.261 \pm  0.002     $    \\
$\alpha$ (rad)                 &  $   2.615 \pm  0.009  $   &  $2.415 \pm  0.005     $    \\
$s_s$                          &  --                        &  $0.079 \pm  0.001     $    \\
$q_s$                          &  --                        &  $0.103 \pm  0.001     $    \\ 
$\phi$ (rad)                   &  --                        &  $1.326 \pm  0.012     $    \\
$ds_s/dt$ (yr$^{-1}$)          &  --                        &  $-0.064 \pm  0.031    $    \\ 
$d\phi/dt$ (deg/day)           &  --                        &  $53.79 \pm  0.30      $    \\
$\rho_1$ (10$^{-3}$)           &  $   3.29  \pm  0.17   $   &  $4.58  \pm  0.05      $    \\ 
$\rho_2$ (10$^{-3}$)           &  $   2.23  \pm  0.79   $   &  $0.10  \pm  0.01      $    \\ 
$q_F$                          &  $   0.71  \pm  0.05   $   &  $0.15  \pm  0.01      $    \\ 
\hline
\end{tabular*}
\tablefoot{ ${\rm HJD}^\prime = {\rm HJD}- 2450000$.  }
\end{table}

\subsection{2L2S interpretation}\label{sec:three-two}

The fact that the two caustic-crossing features can be described by a 2L1S model suggests that 
the other anomaly feature around $t_2$ may be explained with the introduction of an extra source 
(2L2S) or an extra lens component (3L1S) to the 2L1S configuration. In this section, we present 
the analysis conducted under the 2L2S interpretation.

The 2L2S modeling was carried out based on the 2L1S solution. With the initial parameters
obtained from the 2L1S modeling, we searched for a 2L2S solution by testing various trajectories
of the second source considering the location and magnitude of the anomaly around $t_2$.  We first 
check the case in which the position of the source companion with respect to the primary source 
does not vary:  "static 2L2S model".  The model curve of the best-fit static 2L2S solution is 
presented in Figure~\ref{fig:two}, and the lensing parameters are listed in Table~\ref{table:one}.  
The lens-system configuration of the solution is shown in the middle inset of the top panel.  The 
caustic configuration is similar to that of the 2L1S solution except that there are two source 
trajectories of $S_1$ and $S_2$, which are marked by black and blue arrowed lines, respectively.  
According to the solution, the source companion approached the lens slightly earlier than the 
primary source with an impact parameter greater than that of the primary source. The second source 
passed the tip of the upper left caustic cusp, and this produced the anomaly feature around $t_2$.  
Although the static 2L2S model can give rise to a sharp anomaly features appearing in the inner 
region between the caustic spikes, it was found that the residual from the 2L2S model was 
substantial, as shown in the second lower panel of Figure~\ref{fig:two}.

\begin{figure}[t]
\includegraphics[width=\columnwidth]{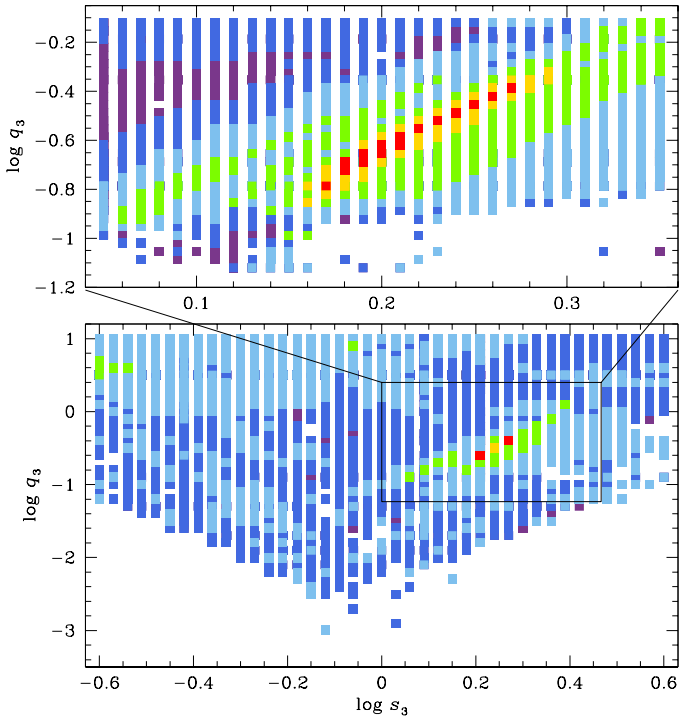}
\caption{
Map of $\Delta\chi^2$ on the $\log~s_3$--$\log~q_3$ parameter plane obtained from the initial 
grid searches for $(s_3, q_3)$ parameters.  Color coding is set to represent points with $\leq 
1n\sigma$ (red), $\leq 2n\sigma$ (yellow), $\leq 3n\sigma$ (green), $\leq 4n\sigma$ (cyan), 
$\leq 5n\sigma$ (blue), and $\leq 6n\sigma$ (purple), where $n=9$.  The upper panel shows the 
map around the minima constructed with denser grids.  
}
\label{fig:three}
\end{figure}

We further check whether the anomalies can be explained by considering the orbital motion of 
the source: "orbital 2L2S model".  We consider the source orbital motion by introducing five 
extra parameters of ($s_s$, $q_s$, $\phi$, $ds_s/dt$, $d\phi/dt$), where $s_s$ and $q_s$ 
represent the normalized separations and mass ratio between the source components, respectively, 
$\phi$ is the orientation angle of $S_2$ with respect to $S_1$, and $(ds_s/dt, d\phi/dt)$ 
represent the change rates of the $s_s$ and $\phi$ induced by the source orbital motion, 
respectively.  It is found that the consideration of the source orbital motion substantially 
improves the fit, by $\Delta\chi^2=275.5$, with respect to the static model.  The model curve 
of the best-fit orbital 2L2S solution is drawn in the top panel of Figure~\ref{fig:two}, 
residual is shown in the panel labeled "2L2S (orbit)", and the lensing parameters are listed 
in Table~\ref{table:one}.  According to the model, the anomaly features around $t_1$ and $t_3$ 
were produced by the primary source as is in the static case, and the feature around $t_2$ was 
produced by the secondary source.  We note that the secondary source crossed the caustic two 
more times at ${\rm HJD}^\prime\sim 9368.7$ ($t_0$) and $\sim 9372.7$, and the former crossing 
is not covered by the data and the latter crossing feature is difficult to identify in the data.
We note that the source radius ratio $\rho_2/\rho_1=1/46$ and the source flux ratio $q_F=0.15$ 
would lead to inconsistent estimates of $\theta_{\rm E}$, which would argue against the physical 
plausibility of this solution. However, we do not pursue this issue because this solution will 
prove to be heavily disfavored by $\chi^2$.  See Sect.~\ref{sec:three-three}.

\begin{figure}[t]
\includegraphics[width=\columnwidth]{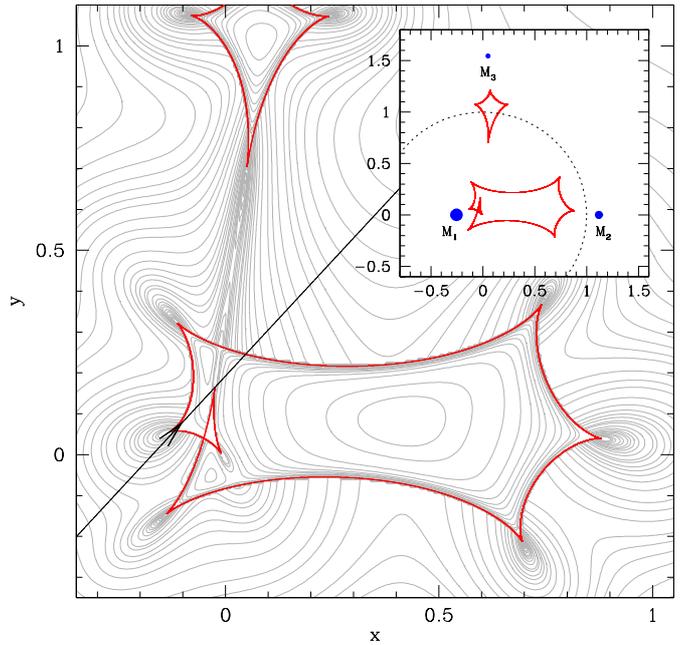}
\caption{
Lens-system configuration of the 3L1S model.  The main panel shows the central anomaly region, 
and the inset displays the whole view of the lens system, where blue dots marked by $M_1$, 
$M_2$, and $M_3$ denote the positions of the individual lens components. Grey curves around 
the caustic represent the equi-magnification contours. The dotted circle centered at the 
origin of the coordinates represents the Einstein ring of the lens system.
}
\label{fig:four}
\end{figure}

\begin{table}[t]
\small
\caption{Lensing parameters of the best-fit 3L1S solution\label{table:two}}
\begin{tabular*}{\columnwidth}{@{\extracolsep{\fill}}lcccc}
\hline\hline
\multicolumn{1}{c}{Parameter}    &
\multicolumn{1}{c}{Value }       \\
\hline
$\chi^2$/dof          &   1399.8/1425               \\
$t_0$ (HJD$^\prime$)  &  $9370.226 \pm 0.061   $    \\
$u_0$                 &  $0.126 \pm 0.008      $    \\
$\te$ (days)          &  $14.74 \pm 0.93       $    \\
$s_2$                 &  $1.386 \pm 0.027      $    \\
$q_2$                 &  $0.526 \pm 0.039      $    \\
$\alpha$ (rad)        &  $2.292 \pm 0.013      $    \\
$s_3$                 &  $1.601 \pm 0.043      $    \\
$q_3$                 &  $0.241 \pm 0.027      $    \\ 
$\psi$ (rad)          &  $1.383 \pm 0.010      $    \\
$\rho$ ($10^{-3}$)    &  $2.50 \pm 0.21        $    \\
\hline
\end{tabular*}
\tablefoot{ ${\rm HJD}^\prime = {\rm HJD}- 2450000$.  }
\end{table}

\subsection{3L1S interpretation}\label{sec:three-three}

We additionally searched for a model under the 3L1S interpretation. Similar to the case of the 
2L2S modeling, the 3L1S modeling was done based on the 2L1S solution, from which we adopted the 
initial parameters of $(t_0, u_0, \te, s_2, q_2, \alpha, \rho)$. We then conducted preliminary 
searches for the parameters related to $M_3$, that is, $(s_3, q_3, \psi)$, via a grid approach, 
and finally refined the solution by letting all parameters vary. Figure~\ref{fig:three} shows 
the $\Delta\chi^2$ map on the $\log~s_3$--$\log~q_3$ parameter plane constructed from the 
preliminary grid search.  The map shows that there exists a unique $\chi^2$ minimum at $(\log~s_3, 
\log~q_3) \sim (0.2, -0.6)$.

It was found that the model identified under the 3L1S interpretation well described all the 
observed anomaly features in the lensing light curve. The lensing parameters of the best-fit 
3L1S solution are listed in Table~\ref{table:two} together with the value of $\chi^2/{\rm dof}$.  
In Figure~\ref{fig:two}, we draw the model curve (solid curve drawn over the data points) and 
residual (presented in the panel labeled "3L1S") from the model. The parameters related to 
$M_2$ and $M_3$ are $(s_2, q_2)\sim (1.4, 0.53)$ and $(s_3, q_3)\sim (1.60, 0.24)$, respectively, 
indicating that the primary of the lens is accompanied by two lower-mass companions lying in 
the vicinity of the Einstein ring. If the separations among the lens components were intrinsic 
(3-dimensional), the three-body lens system would be dynamically unstable, and this indicates 
that one or both companions are likely to be aligned by chance with the primary. We found that 
it was difficult to detect microlens-parallax effects \citep{Gould1992} due to the relatively
 short time scale, $\te\sim 15$~days, of the event compared to the orbital period of Earth.

The lens-system configuration of the 3L1S model is shown in Figure~\ref{fig:four}, in which 
the main panel shows the central anomaly region and the inset shows the whole view of the lens 
system. The coordinates of the configuration is centered at the effective position of the 
$M_1$--$M_2$ pair defined by \citet{Stefano1996} and \citet{An2002}, and the grey curves 
encompassing the caustic represent the equi-magnification contours. One finds that the caustic 
induced by $M_2$ is similar to that of the 2L1S solution, but the 3L1S caustic and the resulting 
magnification pattern around the caustic are distorted by $M_3$, which makes the central part 
of the caustic nested and self-intersected \citep{Gaudi1998, Danek2015a, Danek2015b, Danek2019}. 
According to the 3L1S solution, the anomaly appearing around $t_2$ was produced by the source 
approach close to the tip of the swallow-tail caustic part induced by $M_3$.

\begin{figure}[t]
\includegraphics[width=\columnwidth]{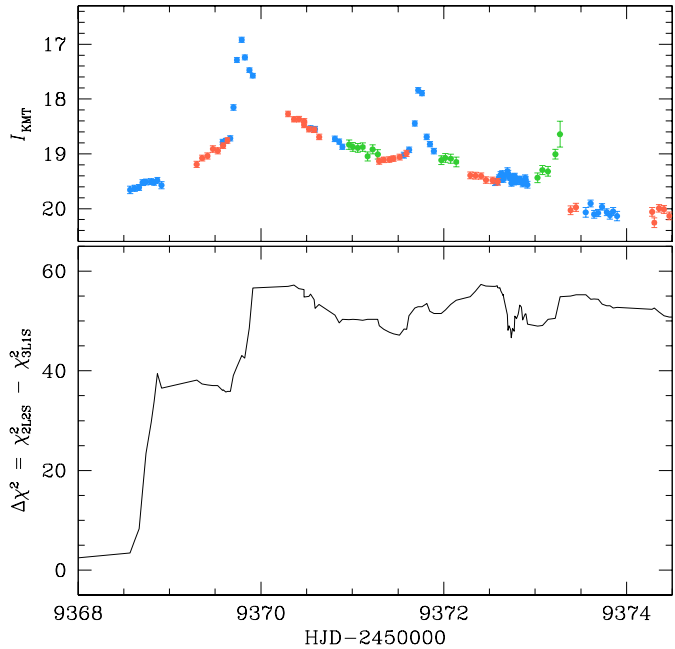}
\caption{
Cumulative distribution of the $\chi^2$ difference, 
$\Delta\chi^2 =\chi^2_{\rm 2L2S}- \chi^2_{\rm 3L1S}$, between the 2L2S (orbital) 
and 3L1S models in the region of the major anomaly.
The light curve in the upper panel is shown to indicate the epochs of 
major $\chi^2$ differences.
}
\label{fig:five}
\end{figure}

From the comparison of the 3L1S and 2L2S (orbital) models, we find that the anomaly features 
in the lensing light curve are better explained by the 3L1S interpretation than by the 2L2S 
interpretation.  In Figure~\ref{fig:five}, we present the cumulative distribution of the 
$\chi^2$ difference, $\Delta\chi^2 =\chi^2_{\rm 2L2S}- \chi^2_{\rm 3L1S}$, between the two 
models in the region of the anomaly for the detailed comparison of the two models.  The major 
differences occur around $t_0$ and $t_1$, at around which $S_2$ and $S_1$ first crossed the 
caustic according to the 2L2S model, respectively.  Besides these regions, the 3L1S model 
well explains the KMTA data point at the epoch of ${\rm HJD}^\prime = 9373.272$, while the 
residual of this data point from the 2L2S model, $\Delta I=0.656$~mag, is very big.  As a 
whole, the 3L1S model yields a better fit than the 2L2S model by $\Delta\chi^2 =53.8$,
despite the fact that the 3L1S model has 4 fewer parameters.

\section{Source star and Einstein radius}\label{sec:four}

We specify the source of the event not only for the full characterization of the event but 
also for the estimation of the angular Einstein radius. The Einstein radius is determined 
from the angular radius of the source, $\theta_*$, by the relation
\begin{equation}
\thetae={\theta_*\over \rho},
\label{eq1}
\end{equation}
where $\theta_*$ is deduced from the source type, and the normalized source radius $\rho$ is 
measured from the modeling by analyzing the caustic-crossing parts of the light curve.

We determined the source type by measuring the extinction- and reddening-corrected color and
brightness, $(V-I, I)_{0,S}$. For this, we utilize the \citet{Yoo2004} method, in which the 
source is located in the instrumental color-magnitude diagram (CMD), and then its color and 
magnitude are calibrated using the centroid of the red giant clump (RGC) as a reference. The 
RGC centroid can be used as a reference for calibration because its de-reddened color and 
magnitude are known.

\begin{figure}[t]
\includegraphics[width=\columnwidth]{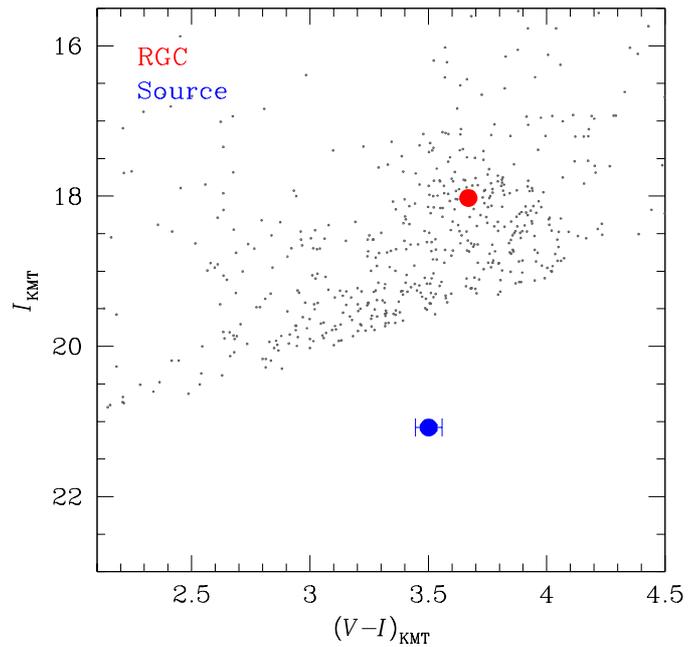}
\caption{
Position of the source (blue dot) with respect to the centroid of red giant clump (RGC, red 
dot) in the instrumental color-magnitude diagram of stars lying in the vicinity of the source.
}
\label{fig:six}
\end{figure}

Figure~\ref{fig:six} shows the position of the source in the instrumental CMD constructed 
from the pyDIA photometry of stars lying around the source.  We note that the CMD shows only 
bright stars due to the severe extinction, $A_I=3.3$,  toward the field.  The measured 
instrumental color and magnitude of the source are
\begin{equation}
(V-I, I)_S = (3.501\pm 0.056, 21.080\pm 0.037), 
\label{eq2}
\end{equation}
where the $I$ and $V$-band magnitudes were estimated by regressing the KMTC light curve data
of the individual passbands processed using the same pyDIA code with respect to the model. By
measuring the offsets in the color and magnitude, $\Delta (V-I, I)$, of the source from the RGC
centroid, with $(V-I, I)_{\rm RGC}=(3.669, 18.025)$, together with the known de-reddened color 
and magnitude of the RGC centroid, $(V-I, I)_{0,{\rm RGC}}=(1.060, 14.377)$ \citep{Bensby2013, 
Nataf2013}, the de-reddened source color and magnitude were measured as
\begin{equation}
\eqalign{
(V-I, I)_{0,S} & = (V-I, I)_{0,{\rm RGC}} + \Delta (V-I, I)  \cr
               & = (0.892\pm 0.056, 17.432\pm 0.037).  \cr
}
\label{eq3}
\end{equation}
The estimated color and magnitude indicate that the source is a subgiant star of an early K 
spectral type.

The angular source radius was deduced from the measured source color and magnitude. For 
this, we first converted $V-I$ color into $V-K$ color using the color-color relation of 
\citet{Bessell1988}, and then estimated the angular source radius using the relation between 
$(V-K, V)$ and $\theta_*$ of \citet{Kervella2004}.  This procedure yielded the angular radii 
of the source and Einstein ring of
\begin{equation}
\theta_* = (1.26 \pm 0.11)~\mu{\rm as}.
\label{eq4}
\end{equation}
and
\begin{equation}
\thetae = 
{\theta_* \over \rho} =
(0.50 \pm 0.06)~{\rm mas}, 
\label{eq5}
\end{equation}
respectively. Together with the measured event time scale, the relative lens-source proper 
motion was estimated as
\begin{equation}
\mu = {\thetae \over \te}  = (12.49 \pm 1.53)~{\rm mas}~{\rm yr}^{-1}.
\label{eq6}
\end{equation}

\section{Physical lens parameters}\label{sec:five}

In this section, we estimate the physical lens parameters of the mass $M$ and distance $\dl$ 
to the lens. These parameters can be sometimes constrained by measuring lensing observables 
of $(\te, \thetae, \pie)$, where $\pie$ denotes the microlens parallax.  The first two 
observables are related to the physical parameters by
\begin{equation}
\te={\thetae\over \mu};\qquad  \thetae=(\kappa M \pi_{\rm rel})^{1/2},
\label{eq7}
\end{equation}
where $\kappa=4G/(c^2{\rm AU})\simeq 8.14~{\rm mas}/M_\odot$ and $\pi_{\rm rel}=\pi_{\rm L}-
\pi_{\rm S}={\rm AU}(1/\dl - 1/\ds)$ represents the relative lens-source parallax.  With the 
additional measurement of the observable $\pie$, the physical parameters can be uniquely 
constrained by the relations \citep{Gould2000}
\begin{equation}
M={\thetae \over \kappa\pie};\qquad \dl = {{\rm au} \over \pie\thetae + \pi_{\rm S}}.
\label{eq8}
\end{equation}
For KMT-2021-BLG-1122, the observables $\te$ and $\thetae$ were securely measured, but 
the other observable $\pie$ could not be measured, and thus we estimate the physical lens 
parameters by conducting a Bayesian analysis using a Galactic model together with the 
constraints provided by the measured observables.

In the first step of the Bayesian analysis, a large number of artificial lensing events were 
produced from a Monte Carlo simulation. For each simulated event, we derived the locations 
of the lens and source and their transverse velocity from a Galactic model, and assigned the 
lens mass from a mass-function model.  In the Monte Carlo simulation, we adopted the Galactic 
model of \citet{Jung2021}, whose detailed description are given therein.  With the simulated 
events, we then constructed the posteriors of the lens mass and distance by giving a weight 
to each simulated event of
\begin{equation}
w_i = \exp(-\chi^2_i/2) ;\qquad
\chi_i^2 = 
{(t_{{\rm E},i} - t_{\rm E})^2 \over [\sigma(\te)]^2} +
{(\theta_{{\rm E},i} - \theta_{\rm E})^2 \over [\sigma(\thetae)]^2},
\label{eq9}
\end{equation}
where $(\te, \thetae)$ and $[\sigma(\te), \sigma(\thetae)]$ represent the measured values 
of the observables and their uncertainties, respectively, and $(t_{{\rm E},i}, \theta_{{\rm E},i})$ 
indicate the observables of each simulated event.

\begin{figure}[t]
\includegraphics[width=\columnwidth]{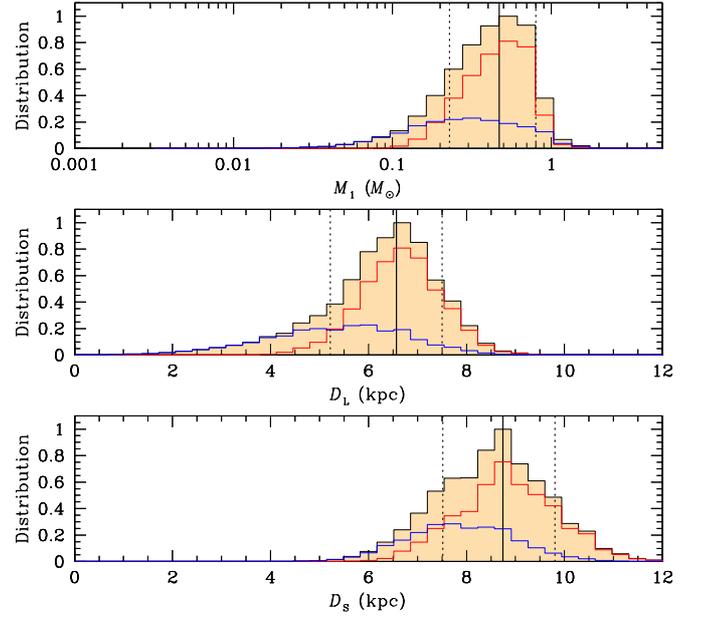}
\caption{
Posteriors of the primary lens mass (top panel) and distance to the lens (middle panel) 
and source (bottom panel). In each panel, the blue and red curves represent the contributions
 by the disk and bulge lens populations, respectively, and the black curve is sum of the two 
lens populations. The solid vertical line indicates the median value, and the two dotted 
lines represent the $1\sigma$ range of the distribution.
}
\label{fig:seven}
\end{figure}

Figure~\ref{fig:seven} shows the posteriors of the primary lens mass (top panel), and distances 
to the lens (middle panel) and source (bottom panel).  We additionally present the $\ds$ 
posterior to show the relative location of the lens and source. For each posterior, we mark 
the contributions by the disk and bulge lens populations with curves drawn in blue and red, 
respectively. The masses of the individual lens components estimated from the Bayesian 
analysis are
\begin{equation}
M_1 = 0.47^{+0.24}_{-0.33}~M_\odot,
\label{eq10}
\end{equation}
\begin{equation}
M_2 = 0.24^{+0.13}_{-0.17}~M_\odot, 
\label{eq11}
\end{equation}
and
\begin{equation}
M_3 = 0.11^{+0.06}_{-0.08}~M_\odot. 
\label{eq12}
\end{equation}
The estimated masses of all the lens components are in the mass regime of M~dwarfs, and 
thus the lens is a triple system composed of three low-mass stars.  The estimated distance 
to the lens system is 
\begin{equation}
\dl = 6.57^{+1.34}_{-0.94}~{\rm kpc}, 
\label{eq13}
\end{equation}
and the 
probabilities for the lens to be in the disk and bulge are 33\% and 67\%, respecively. For 
each parameter, we present the median as a representative value and the lower and upper 
limits are estimated as the 16\% and 84\% ranges of the posterior distribution. The projected 
separations of the secondary and tertiary lens components from the primary are
\begin{equation}
a_{\perp, 2} = s_2\dl \thetae = 3.47^{+0.71}_{-0.50}~{\rm AU},
\label{eq14}
\end{equation}
and
\begin{equation}
a_{\perp, 3} = s_3\dl \thetae = 4.01^{+0.82}_{-0.57}~{\rm AU}, 
\label{eq15}
\end{equation}
respectively.  As noted, the fact that the projected separations $a_{\perp, 2}$ and $a_{\perp, 3}$ 
are similar to each other suggests that one or both lens companions are aligned by chance with 
the primary.

We note that KMT-2021-BLG-1122L is the first triple stellar system detected via microlensing.
In total, there exist 12 confirmed 3L1S events: 
OGLE-2006-BLG-109 \citep{Gaudi2008, Bennett2010}, 
OGLE-2012-BLG-0026 \citep{Han2013}, 
OGLE-2018-BLG-1011 \citep{Han2019}, 
OGLE-2019-BLG-0468 \citep{Han2022-0468},
OGLE-2021-BLG-1077 \citep{Han2022-1077}, 
OGLE-2006-BLG-284 \citep{Bennett2020}, 
OGLE-2007-BLG-349 \citep{Bennett2016},
OGLE-2008-BLG-092 \citep{Poleski2014}, 
OGLE-2013-BLG-0341 \citep{Gould2014}, 
OGLE-2018-BLG-1700 \citep{Han2020-1700}, 
KMT-2019-BLG-1715 \citep{Han2021-1715}, and 
KMT-2020-BLG-0414 \citep{Zang2021}.\footnote{
Besides these confirmed cases, there are 6 candidate triple-lensing events, including
OGLE-2014-BLG-1722 \citep{Suzuki2018}, 
OGLE-2018-BLG-0532 \citep{Ryu2020}, 
KMT-2019-BLG-1953 \citep{Han2020-1953}, 
KMT-2019-BLG-0304 \citep{Han2021-0304}, 
OGLE-2019-BLG-1470 \citep{Kuang2022}, and 
KMT-2021-BLG-0240 \citep{Han2022-0240}.
For these events, 3L1S interpretations yielded best-fit models, but the signals of the 
tertiary lens components were not firmly confirmed either due to the subtlety of the 
signals or the degeneracy with other interpretations.  Under the 3L1S interpretations of 
these events, the lenses belong to either the category of two-planet systems or the category 
of planets in binaries similar to the cases of the confirmed 3L1S events.} Among them, the 
lenses of the first 5 events are two-planet systems, and the lenses of the other 7 events 
are binary-star systems possessing planets.  Therefore, the common trait of the previously 
discovered microlensing triple systems is that at least one of the lens components is a 
planet.  Two planets with similar separations from a host can be dynamically stable if the 
planets are in mean-motion resonance \citep{Madsen2019}.  A planet in a binary system can 
be dynamically stable if the planet orbits one star of a wide binary stellar system or the 
barycenter of a closely-spaced binary system.  Microlensing detection of a triple stellar 
system is difficult because a system with similar masses and intrinsic separations among the 
components would be dynamically unstable.  One way for such a system to be detected via 
microlensing is that the companions of the triple system are closely aligned with the primary 
so that they lie around the Einstein ring of the primary.  This alignment requires projection 
effect of a low probability, and this explains the relative rareness of triple stellar systems 
detected by microlensing.

\section{Summary and conclusion}\label{sec:six}

We presented the analysis of the microlensing event KMT-2021-BLG-1122, which was 
investigated as a part of the project to analyze anomalous lensing events in the
previous KMTNet data with no suggested plausible models. We confirmed that the light 
curve, characterized by three major anomaly features, of the event could not be 
explained with the usual 2L1S or 1L2S interpretations, but it was found that a 2L1S 
solution obtained from the modeling with the exclusion of the data around the second 
anomaly feature could approximately describe the other two caustic-crossing features.

We found that all the anomaly features could be explained by introducing a tertiary 
lens component, and thus the lens is a triple system. The masses of the individual 
lens components estimated from the Bayesian analysis are $(M_1, M_2, M_3)\sim 
(0.47~M_\odot, 0.24~M_\odot, 0.11~M_\odot)$, and the companions are separated in 
projection from the primary by $(a_{\perp,2}, a_{\perp,3})\sim (3.5, 4.0)$~AU.  The 
estimated masses of all the lens components are in the mass regime of M dwarfs, and 
thus the lens is a triple system composed of three low-mass stars.  This is the first 
triple stellar-lens system detected via microlensing.

\begin{acknowledgements}
Work by C.H. was supported by the grants of National Research Foundation of Korea 
(2020R1A4A2002885 and 2019R1A2C2085965).
J.C.Y. acknowledges the support from NSF Grant No. AST-2108414.
W.Z. and H.Y. acknowledge the support by the National Science Foundation of
China (Grant No. 12133005). 
Y.S. acknowledges support from BSF Grant No. 2020740.
This research has made use of the KMTNet system operated by the Korea Astronomy and Space 
Science Institute (KASI) and the data were obtained at three host sites of CTIO in Chile, 
SAAO in South Africa, and SSO in Australia.
\end{acknowledgements}


\begin{thebibliography}{}
\bibitem[Albrow(2017)]{Albrow2017} Albrow, M.\ 2017, MichaelDAlbrow/pyDIA: Initial Release on Github,Versionv1.0.0, Zenodo, doi:10.5281/zenodo.268049
\bibitem[Albrow et al.(2009)]{Albrow2009} Albrow, M., Horne, K., Bramich, D.~M., et al.\ 2009, \mnras, 397, 2099
\bibitem[An \& Han(2002)]{An2002} An, J. H., \& Han, C. 2002, \apj, 573, 351
\bibitem[Beaulieu et al.(2016)]{Beaulieu2016} Beaulieu, J.-P., Bennett, D.~P., Batista, V., et al.\ 2016, \apj, 824, 83
\bibitem[Bennett \& Rhie(1996)]{Bennett1996} Bennett, D.~P., \& Rhie, S.~H.\ 1996, \apj, 472, 660
\bibitem[Bennett et al.(2010)]{Bennett2010} Bennett, D.~P., Rhie, S.~H., Nikolaev, S., et al.\ 2010, \apj, 713, 837
\bibitem[Bennett et al.(2016)]{Bennett2016} Bennett, D. P., Rhie, S. H., Udalski, A., et al. 2016, \aj, 152, 125 
\bibitem[Bennett et al.(2020)]{Bennett2020} Bennett, D.~P., Udalski, A., Bond, I. A., et al.\ 2020, \aj, 160, 72
\bibitem[Bensby et al.(2013)]{Bensby2013} Bensby, T., Yee, J.~C., Feltzing, S., et al.\ 2013, \aap, 549, A147
\bibitem[Bond et al.(2001)]{Bond2001} Bond, I. A., Abe, F., Dodd, R. J., et al. 2001, \mnras, 327, 868 
\bibitem[Bessell \& Brett(1988)]{Bessell1988} Bessell, M.~S., \& Brett, J. M. 1988, \pasp, 100, 1134
\bibitem[Dan\u{e}k \& Heyrovsk\'y(2015a)]{Danek2015a} Dan\u{e}k, K., \& Heyrovsk\'y, D. 2015a, \apj, 806, 63
\bibitem[Dan\u{e}k \& Heyrovsk\'y(2015b)]{Danek2015b} Dan\u{e}k, K., \& Heyrovsk\'y, D. 2015b, \apj, 806, 99
\bibitem[Dan\u{e}k \& Heyrovsk\'y(2019)]{Danek2019}   Dan\u{e}k, K., \& Heyrovsk\'y, D. 2019, \apj, 880, 72
\bibitem[Di Stefano \& Mao(1996)]{Stefano1996} Di Stefano, R., \& Mao, S. 1996, \apj, 457, 93
\bibitem[Dong et al.(2008)]{Dong2006} Dong, S., DePoy, D. L., Gaudi, B. S., et al. 2006, \apj, 642, 842
\bibitem[Gaudi et al.(2008)]{Gaudi2008} Gaudi, B. S., Bennett, D. P., Udalski, A., et al.\ 2008, Science, 319, 927
\bibitem[Gaudi et al.(1998)]{Gaudi1998} Gaudi, B. S., Naber, R. M., \& Sackett, P. D.\ 1998, \apj, 502, L33
\bibitem[Gould(1992)]{Gould1992} Gould, A. 1992, \apj, 392, 442
\bibitem[Gould(2000)]{Gould2000} Gould, A. 1992, \apj, 542, 785
\bibitem[Gould et al.(2014)]{Gould2014} Gould, A., Udalski, A., Shin, I. -G., et al. 2014, Science, 345, 46
\bibitem[Griest \& Hu(1993)]{Griest1993} Griest, K., \& Hu, W. 1993, \apj, 407, 440
\bibitem[Han et al.(2019)]{Han2019}        Han, C., Bennett, D.~P., Udalski, A., et al.\ 2019, \aj, 158, 114
\bibitem[Han \& Gould(1997)]{Han1997}      Han, C., \& Gould, A.\ 1997, \apj, 480, 196
\bibitem[Han et al.(2022a)]{Han2022-1077}  Han, C., Gould, A., Bond, I. A., et al.\ 2022a, \aap, 662, A70 
\bibitem[Han et al.(2022a)]{Han2020-1953}  Han, C., Kim, D., Jung, Y. K. 2020a, \aj, 160, 17
\bibitem[Han et al.(2022b)]{Han2022-0240}  Han, C., Kim, D., Yang, H. 2022b, \aap, 664, 114
\bibitem[Han et al.(2023)]{Han2023}        Han, C., Lee, C.-U., Gould, A., et al. 2023, \aap, in press 
\bibitem[Han et al.(2020b)]{Han2020-1700}  Han, C., Lee, C.-U., Udalski, A., et al.\ 2020b, \aj, 159, 48                       
\bibitem[Han et al.(2022c)]{Han2022-0890}  Han, C., Ryu, Y.-H., Shin, I.-G., et al. 2022c, \aap, 667, A64
\bibitem[Han et al.(2013)]{Han2013}        Han, C., Udalski, A., Choi, J.-Y., et al.\ 2013, \apj, 762, L28
\bibitem[Han et al.(2021a)]{Han2021-1715}  Han, C., Udalski, A., Kim, D., et al. 2021a, \aj, 161, 270 
\bibitem[Han et al.(2021b)]{Han2021-0304}  Han, C., Udalski, A., Lee, C.-U. 2021b, \aj, 162, 203 
\bibitem[Han et al.(2022d)]{Han2022-0468}  Han, C., Udalski, A., Lee, C.-U. 2022d, \aap, 658, A93 
\bibitem[Hwang et al.(2018)]{Hwang2018} Hwang, K. -H., Udalski, A., Shvartzvald, Y., et al. 2018, \aj, 155, 20
\bibitem[Jung et al.(2021)]{Jung2021}  Jung, Y. K., Han, C., Udalski, A., et al. 2021, \aj, 161, 293
\bibitem[Kervella et al.(2004)]{Kervella2004} Kervella, P., Th\'evenin, F., Di Folco, E., \& S\'egransan, D.\ 2004, \aap, 426, 29
\bibitem[Kim et al.(2018a)]{Kim2018a} Kim, D.-J., Kim, H.-W., Hwang, K.-H., et al. 2018a, \aj, 155, 76
\bibitem[Kim et al.(2018b)]{Kim2018b} Kim, H.-W., Hwang, K.-H., Shvartzvald, Y., et al. 2018b, arXiv:1806.07545
\bibitem[Kim et al.(2016)]{Kim2016} Kim, S.-L., Lee, C.-U., Park, B.-G., et al.\ 2016, JKAS, 49, 37
\bibitem[Kuang et al.(2022)]{Kuang2022} Kuang, R., Zang, W., Jung, Y. K., et al. 2022, \mnras, 516, 1704
\bibitem[Madsen \& Zhu(2019)]{Madsen2019} Madsen, S., \& Zhu, W. 2019, \apj, 878, L29
\bibitem[Mao \& Paczy\'nski(1991)]{Mao1991} Mao, S., \& Paczy\'nski, B., 1991, \apj, 374, L37 
\bibitem[Nataf et al.(2013)]{Nataf2013} Nataf, D.~M., Gould, A., Fouqu\'e, P., et al.\ 2013, \apj, 769, 88
\bibitem[Paczy\'nski(1986)]{Paczynski1986} Paczy\'nski, B. 1986, \apj, 301, 503 
\bibitem[Poleski et al.(2014)]{Poleski2014} Poleski, R., Skowron, J., Udalski, A., et al.\ 2014, \apj, 795, 42
\bibitem[Ryu et al.(2020)]{Ryu2020} Ryu, Y.-H., Udalski, A., Yee, J. C., et al.\ 2020, \aj, 160, 183
\bibitem[Schneider \& Weiss(1986)]{Schneider1986} Schneider, P., \& Weiss, A. 1986, \aap, 164, 237
\bibitem[Schneider \& Weiss(1987)]{Schneider1987} Schneider, P., \& Weiss, A. 1987, \aap, 171, 49
\bibitem[Suzuki et al.(2018)]{Suzuki2018} Suzuki, D., Bennett, D. P., Udalski, A., et al. 2018, \aj, 155, 263
\bibitem[Udalski et al.(1994)]{Udalski1994} Udalski, A., Szyma\'nski, M., Ka{\l}u\.zny, J., et al. 1994, \actaa, 44, 1
\bibitem[Udalski et al.(2015)]{Udalski2015} Udalski, A., Szyma\'nski, M.~K., \& Szyma\'nski, G.\ 2015, \actaa, 65, 1
\bibitem[Yee et al.(2012)]{Yee2012} Yee, J.~C., Shvartzvald, Y., Gal-Yam, A., et al.\ 2012, \apj, 755, 102
\bibitem[Yoo et al.(2004)]{Yoo2004} Yoo, J., DePoy, D.~L., Gal-Yam, A., et al.\ 2004, \apj, 603, 139
\bibitem[Zang et al.(2021)]{Zang2021} Zang, W., Han, C., Kondo, I., et al. 2021, Res. Astron. Astrophys., 21, 239

\vspace*{\fill}
\end{thebibliography}
\end{document}